\documentclass[twocolumn,pre,superscriptaddress,amsmath,amssymb]{revtex4}
\usepackage{latexsym}
\usepackage{graphicx}
\usepackage{epsfig,subfigure}
\begin{document}
%%%%%%%%%%%%%%%%%%%%%%%%%%%%%%%%%%%%%%%%%%%%%%%%%%%%%%%%%%%%%%%%
\title{Computer simulations of history of life: speciation, emergence of complex species from simpler organisms, and extinctions
 }

\author{Debashish Chowdhury}
\affiliation{Department of Physics, Indian Institute of Technology, Kanpur 208016, India.}

\author{Dietrich Stauffer}
\affiliation{Institute for Theoretical Physics, Cologne University, D-50923 K\"
oln, Germany.}

\begin{abstract}

We propose a generic model of eco-systems, with a {\it hierarchical} 
food web structure. In our computer simulations we let the eco-system 
evolve continuously for so long that that we can monitor extinctions 
as well as speciations over geological time scales. {\it Speciation} 
leads not only to horizontal diversification of species at any given 
trophic level but also to vertical bio-diversity that accounts for the 
emergence of complex species from simpler forms of life. We find that 
five or six trophic levels appear as the eco-system evolves for 
sufficiently long time, starting initially from just one single level. 
Moreover, the time intervals between the successive collections of 
ecological data is so short that we could also study ``micro''-evolution 
of the eco-system, i.e., the birth, ageing and death of individual 
organisms.

\end{abstract}


\maketitle

\bigskip 

\section{Introduction}
During last one decade, theoretical research on co-evolution of species 
in eco-systems and the statistics of extinctions have been strongly 
infleunced by the pioneering interdisciplinary works of Per Bak and his 
collaborators \cite{bak,nature,boettcher,tree}. In the same spirit, we 
address some fundamental questions of evolutionary ecology from the 
perspective of statistical physics.

How did the higher species emerge in eco-systems inhabited initially 
only by primitive forms of life, like bacteria and plankton 
\cite{maynard,mayr2001,carroll}?
The available record of the history of life, written on stone in the 
form of fossils, is incomplete and ambiguous. An alternative enterprise 
seeks to recreate the evolution on a computer by simulating theoretical 
models \cite{drosselrev,newmanrev}. In this paper we propose 
a theoretical model that not only addresses the question raised above 
but also provides a versatile conceptual tool for studying evolutionary 
ecology. In particular, it describes both the ``macro''-evolutionary 
processes (e.g., origin, evolution and extinction of species) as well as 
``micro''-evolution, (e.g., age-distribution in the population of a 
species, mortality rates, etc.).  

If watched over a short period of time, the dynamics of the eco-system 
appears to be dominated by birth and death of the individual organisms 
as well as by the prey-predator interactions. However, over longer period 
of time, one would see not only extinction of some species but also the 
appearance of new ones. Besides, in many situations, macro-evolutionary 
changes occur at rates that are comparable to those of the ecological 
processes \cite{thompson98,fussmann}. The artificial separation 
of this process into ``ecological'' time scales and ``geological'' time 
scales \cite{khibnik} has been made in many earlier 
theoretical works only for the convenience of modelling. 

The ``ecological'' models, that describe population dynamics in detail 
using, for example, the Lotka-Volterra equations \cite{goel,sigmund} 
usually ignore the slow macro-evolutionary 
changes in the eco-system; hardly any effects of these would be 
observable before the computer simulations would run out of computer 
time \cite{murray}. On the other hand, in order to simulate the 
billion-year old history of life on earth with a computer, the 
elementary time steps in ``evolutionary'' models have to correspond to 
thousands of years, if not millions; consequently, the finer details 
of the ecological processes over shorter periods of time  cannot be 
accounted for by these models in any explicit manner \cite{kauffman,bak}. 
Limitations of these approaches are well known \cite{caldarelli,higgs,quince}.
Moreover, most of the recent computer models of 
ageing \cite{stauffer} focus attention on only one isolated species 
and, therefore, cannot capture macro-evolutionary phenomena like, for 
example, extinctions which depend crucially on the prey-predator 
interactions.

We wish to develop one single theoretical model which would be able to 
describe the entire 
dynamics of an eco-system since the first appearance of life in it up 
till now and in as much detail as possible. This dream has now come  
closer to reality, mainly because of the availability of fast computers 
\cite{hall,collobiano,chowetal,chowstau,rikvold,cebrat}. 
It has become feasible now to carry out computer simulations (in-silico 
experiments) of eco-system models where, each time step would correspond 
to typical times for ``micro''-evolution while each of the simulations 
is run long enough to capture ``macro''-evolution. 

The prey-predator relations in any eco-system are usually described 
graphically in terms of food webs \cite{pimm,polis,mckane,cohen90}.  
More precisely, a food web 
is a directed graph where each node is labelled by a species' name and 
each directed link indicates the direction of flow of nutrient (i.e., 
{\it from} a prey {\it to} one of its predators). We incorporate in our 
model the hierarchical organization of the species at different trophic 
levels of the food web. 

In real eco-systems, the food web is a slowly evolving dynamic network. 
For example, species are known to change their food habits \cite{thompson99}. 
These changes in diets may be caused by scarcity of the normal 
food and abundance of alternative food resources. Moreover, higher 
organisms appear through speciation in an eco-system that initially 
began with only simple forms of life. These not only occupy new trophic 
levels but also introduce new prey-predator interactions with the 
existing species. Therefore, it is also desirable that these {\it 
self-organizing} features of natural eco-systems should be reproduced, 
at least qualitatively, by the theoretical models.

The aim of this paper is to propose a model that would capture the  
desirable features of eco-systems outlined above.
Higgs, McKane and collaborators \cite{caldarelli,higgs} have developed 
a model, called the Webworld model, which was aimed at linking the 
ecological modeling of food web architecture with the evolutionary 
modeling of speciation and extinction. The spirit of our model is very 
similar although the details of the mathematical formulation of the 
two models are quite different. 

\section{The model}

We model the eco-sytem as a dynamic {\it hierarchical} network.
The ``micro''-evolution, i.e., the birth, growth (ageing) and natural 
death of the individual organisms, in our model is captured by the 
intra-node dynamics. The ``macro``-evolution, e.g., adaptive 
co-evolution of the species, is incorporated in the same model through a 
slower evolution of the network itself over longer time scales. 
Moreover, as the model eco-system evolves with time, extinction of 
species is indicated by vanishing of the corresponding population; 
thus, the number of species and the trophic levels in the model 
eco-system can fluctuate with time. Furthermore, the natural process of 
speciation is implemented by allowing re-occupation of vacant nodes by 
mutated versions of non-extinct species. 

\subsection{Architecture of the network}

Each node of the network represents a niche that can be occupied by at 
most one species at a time. 
The number of nodes in the trophic level ${\ell}$ is $m^{\ell}$ where $m$ is a 
positive integer. We assume only one single species at the highest 
level $\ell = 1$. The allowed range of ${\ell}$ is 
$1 \leq {\ell} \leq {\ell}_{max}(t)$, where ${\ell}_{max}(t)$ is a 
time-dependent number in our new model. In other words, in contrast to 
all cited earlier models, the numerical value of $\ell_{max}$ in our new 
model is not put in by hand, but is an {\it emergent property} of the 
eco-system.

\subsection{Prey-predator interactions and intra-species competitions}

The prey-predator interaction between two species that occupy the 
nodes $i$ and $k$ at two adjacent trophic levels is represented by 
$J_{ik}$; the three possible values of $J_{ik}$ are $\pm 1$ and $0$. 
The sign of $J_{ik}$ indicates the direction of trophic flow, i.e. 
{\it from the lower to the higher} level. $J_{ik}$ is $+1$ if $i$ 
eats $k$ and it is $-1$ if $k$ eats $i$. If there is no prey-predator 
relation between the two species $i$ and $k$, we must have $J_{ik} = 0$. 
Although there is no direct interaction between species at the same 
trophic level in our model, they can compete, albeit indirectly, with 
each other for the same food resources available in the form of prey 
at the next lower trophic level.

We now argue that the elements of the matrix $J$ account not only for 
the {\it inter}-species interactions but also for the {\it intra}-species 
interactions arising from the competition of individual organisms for 
the same food resources. Let $S_i^+$ be the number of all prey individuals 
for species $i$ on the lower trophic level, and $S_i^-$ be $m$ times the 
number of all predator individuals on the higher trophic level. Since 
we assume that a predator eats $m$ prey per time interval, $S_i^+$ gives 
the amount of total food available for species $i$, and $S_i^-$ is the 
total contribution of species $i$ to the pool of food required for all 
the predators on the higher level. If the available food $S_i^+$  is 
less than the requirement, then some organisms of the species $i$ will 
die of {\it starvation}, even if none of them is killed by any predator. 

The {\it intra}-species competition among the organisms 
of the same species for limited availability of resources, other than 
food, imposes an upper limit $n_{max}$ of the allowed population of 
each species; $n_{max}$ is time-independent parameter in the model. 
Thus, the total number of organisms $n(t)$ at time $t$ is given by 
$n(t) = \sum_{i=1}^{N(t)} n_i(t)$. 

If $n_i-S_i^+$ is larger than $S_i^-$ then food shortage will be the 
dominant cause of premature death of a fraction of the existing 
population of the species $i$. On the other hand, if $S_i^- > n_i-S_i^+$, 
then a fraction of the existing population will be wiped out primarily 
by the predators. 

In order to capture the {\it starvation deaths and killing by the 
predators}, in addition to the natural death due to ageing, 
a reduction of the population by
\begin{equation}
C ~~\max(S_i^-,~n_i- S_i^{+})
\label{eq-kill}
\end{equation}
is implemented at every time step, where $n_i$ is the population of 
the species $i$ that survives after the natural death. $C$ is a constant
of proportionality. If this leads to $n_i \leq 0$, species $i$ becomes 
extinct.

We assume that the simplest species occupying the lowest trophic 
level always get enough resources that neither natural death nor  
predators can affect their population.

\subsection{Collective characteristics of species}

An arbitrary species $i$ 
is {\it collectively} characterized by \cite{chowetal}:\\
(i) the {\it minimum reproduction age} $X_{rep}(i)$,\\
(ii) the {\it birth rate} $M(i)$,\\
(iii) the {\it maximum possible age} $X_{max}(i)$ 
that depends only on the trophic level occupied by the species. \\
An individual of the species $i$ can reproduce only
after attaining the age $X_{rep}(i)$. Whenever an organism of
this species gives birth to offsprings, $M(i)$ of
these are born simultaneously. None of the individuals of this
species can live longer than $X_{max}(i)$,
even if an individual manages to escape its predators or starvation.

\subsection{Mutations} 

With probability $p_{mut}$ per unit time, each of the species randomly 
increases or decreases, with equal probability, their $X_{rep}$ and $M$ 
by unity. ($X_{rep}$ is restricted to remain in the interval from $1$ 
to $X_{max}$, and $M > 0$.) Moreover, with the same probability $p_{mut}$ 
per unit time, they also re-adjust one of the links $J$ from prey and 
one of the links $J$ to predators \cite{sole}. If the link 
$J_{ij}$ to the species $i$ from a {\it higher} level species $j$ is 
non-zero, it is assigned a new value of $J_{ij} = J_{ji} = 0$. On the 
other hand, if the link $J_{ik}$ to a species $i$ from a {\it lower} 
species $k$ is zero, the new values assigned are $J_{ik} = 1, J_{ki} = -1$. 
These re-adjustments of the incoming and outgoing (in the sense of 
nutrient flow) interactions are intended to capture the facts that each 
species tries to minimize predators but look for new food resources.

\subsection{Speciation} The niches (nodes) left empty because of 
extinction are re-filled by new species, with probability $p_{sp}$ 
per unit time \cite{orr}. All the simultaneously re-filled 
nodes in a trophic level of the network originate from {\it one common 
ancestor} which is picked up randomly from among the non-extinct species 
at the same trophic level. All the interactions $J$ of the new species 
are identical to those of their common ancestor. The characteristic 
parameters $X_{rep}$, $M$ of each of the new species differ randomly 
by $\pm 1$ from the corresponding parameters for their ancestor. 

However, occasionally, all the niches at a level may lie vacant. 
Under such circumstances, all these vacant nodes are to be filled 
by a mutant of the non-extinct species occupying the closest {\it 
lower} populated level. As stated above, the lowest level, that is 
populated by the simplest species, never goes extinct; the possible 
ageing of the species at the lowest level \cite{ackermann}  
is not relevant here. All the 
individual organisms of the new species are assumed to be newborn 
babies that begin ageing with time just like the other species.  
Since space does not enter explicitly in our model, it does 
not distinguish between sympatric and allopatric speciation \cite{dieckmann}.

\subsection{Emergence of new trophic level} 
In order to understand why the total number of 
trophic levels in food webs usually lie between $4$ and $6$, we 
allowed adding a new trophic level to the food web, with a small 
probability $p_{lev}$ per unit time, provided the total bio-mass 
distributed over all the levels (including the new one) does not 
exceed the total bio-mass available in the eco-system. This step is 
motivated by the fact that real ecosystems can exhibit growing 
bio-diversity over sufficiently long period of time. Increase of the 
number trophic level means the diversification at the erstwhile 
topmost level as well as all the lower levels and the emergence of 
yet another dominating species that occupies the new highest level.
The total number of levels, which determines the lengths of the food 
chains, depends on several factors, including the available bio-mass 
\cite{post}. 

\subsection{Birth and natural death of organisms}

At each time step, each individual organism $\alpha$ of the species 
$i$ gives birth {\it asexually} to $M(i)$ offsprings with a probability 
$p_b(i,\alpha)$. We also assume the {\it time-dependent} probability 
$p_b(i,\alpha)$ 
is a product of two factors. One of these two factors decreases linearly 
with age, from unity, attainable at the minimum reproduction age, to 
zero at the maximum lifespan. The other factor is a standard 
Verhulst factor $ 1 - n_i/n_{max}$ which takes into account the fact 
that the eco-system can support only a maximum of $n_{max}$ individual 
organisms of each species. Thus, $p_b(i,\alpha)$ is equal to the Verhulst 
factor at $X = X_{rep}$. 

Each individual organism, irrespective of its age, can meet its 
natural death. However, the probability $p_d$ of this natural death 
depends on the age of the individual. In order to mimic age-independent 
constant mortality rate in childhood, we assume the probability $p_d$ of 
``natural'' death (due to ageing) to be a constant 
$p_d = \exp[-r(X_{max}- X_{rep})/M] $, (with a small fraction $r$), so 
long as $X < X_{rep}$. However, for $X > X_{rep}$, 
the probability of natural death is assumed to increase following the 
Gompertz law $p_d = \exp[-r(X_{max}- X)/M] $. 
Note that, for a given $X_{max}$ and $X_{rep}$,
the larger is the $M$ the higher is the $p_d$ for any age $X$.
Therefore, in order maximize reproductive success, each species has a 
tendency to increase $M$ for giving birth to larger number of offsprings 
whereas the higher mortality for higher $M$ opposes this tendency 
\cite{ghalambor}. However, even with a constant $p_d = 0.1$ 
we found qualitatively similar results.

\subsection{Summary of the dynamics of the eco-system}

The state of the system is updated in discrete time steps where each 
step consists of a sequence of six stages:\\ 

\noindent {\it I- Birth} 

\noindent {\it II- Natural death} 

\noindent {\it III- Mutation} 

\noindent {\it IV- Starvation death and killing by prey} 

\noindent {\it V- Speciation} 

\noindent {\it VI- Emergence of new trophic level}

In all our simulations we began with random initial condition, except 
for $M = 1$ for all species, mostly with only {\it three} levels in 
the food web, and let the eco-system evolve for $T_{w}$ time steps 
before we started collecting ecological and evolutionary data from it; 
these data were collected for the subsequent $T = 5 T_w$ time steps 
where the longest runs were for $T = 10^8$. 
We have not observed any qualitative differences 
in the data  for $n_{max} = 100$ and $n_{max} = 1000$, keeping all the 
other parameters same. Most of  our simulations were carried out with 
$m = 2$, as we did not observe qualitative differences between the data 
for $m = 2$ and $m = 3, 4$ in test runs. The maximum lifespans in the 
levels were assumed to be $X_{max} = 100, 71, 50, 35, 25,....$ starting 
from the highest level.

\section{Results} 

\subsection{Lifetime distributions} 

\begin{figure}[tb]
\begin{center}
\includegraphics[angle=-90,width=0.9\columnwidth]{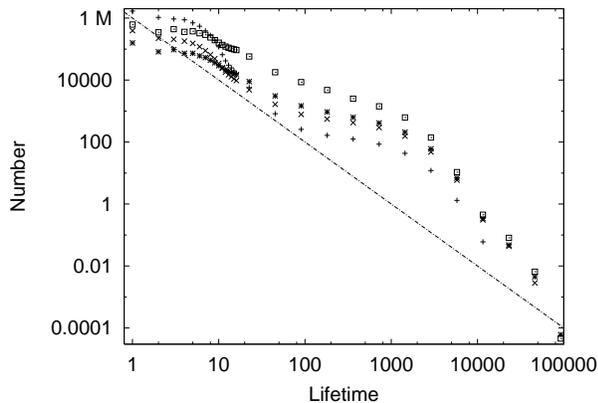}
\end{center}
\caption{Log-log plots of the distributions of the lifetimes of the
species in an eco-system with the total biomass 
of $4 n_{max}$. The symbols $+$, $\times$, $\ast$ and $\square$ 
correspond, respectively, to $10^4$, $10^5$, $10^6$ and $10^8$ 
time steps. Each of the data points have been obtained by averaging 
over $6400$ samples ($+$), $640$ samples ($\times$), $64$ samples 
($\ast$) and $1$ sample ($\square$). The line with slope $-2$ corresponds 
to a power law distribution that has been predicted by many theories.
The common parameters for all the plots are
$m=2$, $n_{max} = 100$, $p_{sp} = 0.1, p_{mut} = 0.001, p_{lev} = 0.0001$,
$C = 0.2$, $r = 0.05$. 
}
\label{fig-1}
\end{figure}

Several theories, based on extremely simple models, claim that the 
distribution of the lifetimes of the species should follow a power 
law with a slope of $-2$ on the log-log plot. The distributions of 
the lifetimes of the species in our model are shown in fig.\ref{fig-1} 
for a few different sets of parameter values. Although our data do 
not rule out an approximate power law over limited regime of 
lifetimes, one single power law over the entire range of lifetimes 
seems impossible.

%%%%%%%%%%%%%%%%%%%%%%%%%%%%%%%%%%%%%%%%%%%%%%%%%%%%%%%%%%%%%%%%%%%%
%%%%%%%%%%%%%%%%%%%%%%%%%%%%%%%%%%%%%%%%%%%%%%%%%%%%%%%%%%%%%%%%%%%%

%%%%%%%%%%%%%%%%%%%%%%%%%%%%%%%%%%%%%%%%%%%%%%%%%%%%%%%%%%%%%%%%%%%%
\begin{figure}[tb]
\begin{center}
\includegraphics[angle=-90,width=0.9\columnwidth]{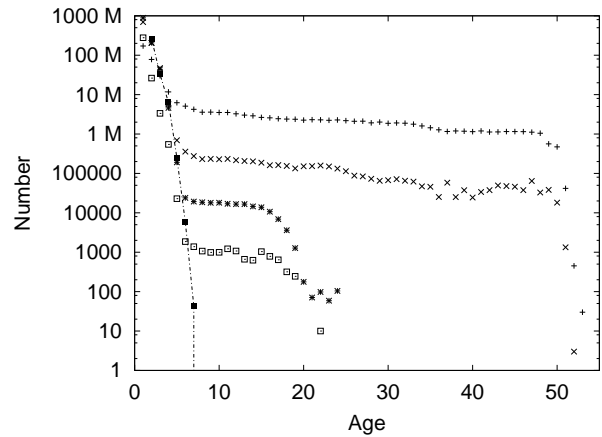}
\end{center}
\caption{Semi-log plots of the distributions of the minimum 
reproductive age $X_{rep}$ of the species. 
The symbols $+$, $\times$, $\ast$, $\square$ and $\blacksquare$
correspond, respectively, to $10^4$, $10^5$, $10^6$, $10^7$ and $10^8$ 
time steps. Each of the data points have been obtained by averaging 
over $6400$ samples ($+$), $640$ samples ($\times$), $64$ samples 
($\ast$) and $1$ sample ($\square$ and $\blacksquare$). 
The values of the common parameters for all the plots are
identical to those used in fig.\ref{fig-1}. 
}
\label{fig-2}
\end{figure}
%%%%%%%%%%%%%%%%%%%%%%%%%%%%%%%%%%%%%%%%%%%%%%%%%%%%%%%%%%%%%%%%%%%%

\subsection{Distribution of minimum reproductive ages} 

In fig.\ref{fig-2} we plot the distributions of the minimum reproductive 
age $X_{rep}$ of the species for several different sets of values of 
the model parameters. Although over relatively short time scales of 
observation this distribution appears quite broad it narrows down with 
evolution and the non-zero values of this distribution correspond to 
reasonable values of age.

\subsection{The number of trophic levels}

\begin{figure}[tb]
\begin{center}
\includegraphics[angle=-90,width=0.9\columnwidth]{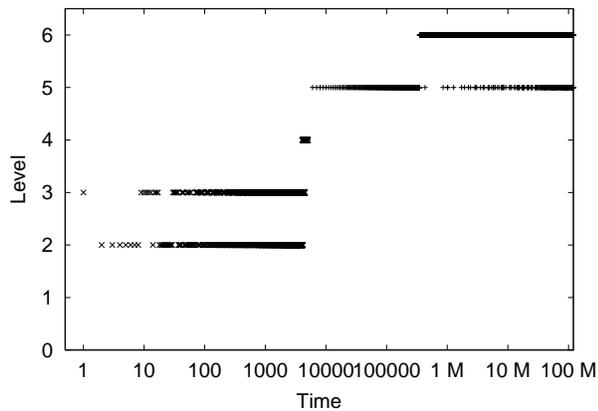}
\end{center}
\caption{Semi-log plot of the number ${\cal N}(t)$ of the trophic 
levels with at least one non-extinct species plotted against time $t$ 
in one single run. The values of the parameters are identical to 
those used in fig.\ref{fig-1}.
}
\label{fig-3}
\end{figure}

Due to the randomness in the evolutionary process, 
occassionally, all of the niches in a level (except the lowest one) 
may lie vacant. We have monitored $\ell_{max}(t)$ and also  
${\cal N}(t)$, the number of those levels at time $t$ in which at 
least one niche is occupied by a non-extinct species. In 
fig.\ref{fig-3} we plot ${\cal N}$ as a function of time for one 
single run. This clearly shows how, over geological time scales, 
$\ell_{max}$ reaches $6$. In this run, the sixth level (the highest 
one) emerges after $10^5$ time steps. It also demonstrates that at 
all stages of evolution, the number ${\cal N}(t)$ keeps fluctuating. 
During the very late stages, ${\cal N}$ keeps fluctuating between 
$5$ and $6$, although ${\cal N}$ is more often $6$ than $5$ for all 
times beyond $10^6$. The ratio of occurrences of {\it six} levels 
and {\it five} levels in the eco-system stabilized only after 
$10^7$ time steps.

\begin{figure}[tb]
\begin{center}
\includegraphics[angle=-90,width=0.9\columnwidth]{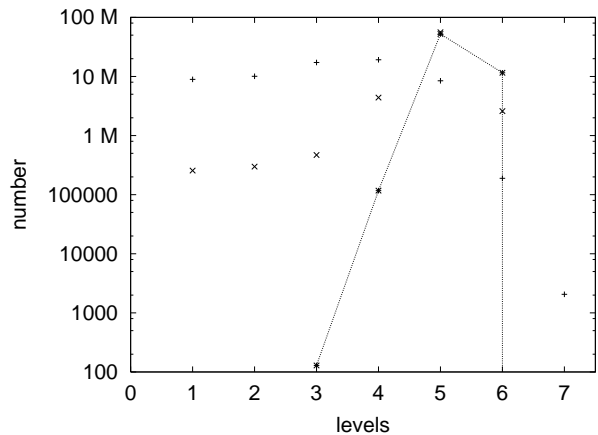}
\end{center}
\caption{Distribution (histogram) of the number ${\cal N}$ of the 
trophic levels with at least one non-extinct species. Identical 
symbols in this figure and in fig.\ref{fig-1} correspond to 
the same runs.  The data for the longest time steps 
are connected by lines to emphasize the shrinking trend of the 
distribution with time.
}
\label{fig-4}
\end{figure}

We have also computed the distributions 
(histograms) of ${\cal N}$ by averaging the data over large 
number of runs. As shown in fig.\ref{fig-4}, the distribution 
becomes narrower for longer runs and the trend indicates than in 
the extreme long time limit would be sharply peaked around one 
single value of ${\cal N}$, as indicated by the fig.\ref{fig-3}.

%%%%%%%%%%%%%%%%%%%%%%%%%%%%%%%%%%%%%%%%%%%%%%%%%%%%%%%%%%%%%%%%%%%%
%%%%%%%%%%%%%%%%%%%%%%%%%%%%%%%%%%%%%%%%%%%%%%%%%%%%%%%%%%%%%%%%%%%%

%%%%%%%%%%%%%%%%%%%%%%%%%%%%%%%%%%%%%%%%%%%%%%%%%%%%%%%%%%%%%%%%%%%%
%%%%%%%%%%%%%%%%%%%%%%%%%%%%%%%%%%%%%%%%%%%%%%%%%%%%%%%%%%%%%%%%%%%%

\section{Summary and conclusion}

In this paper we have introduced a theoretical model of eco-systems 
with a generic hierarchical trophic level structure. Because of data 
collection at suffciently short intervals, we have been able to 
monitor the ecological phenomena like, for example, birth, ageing 
and death of individual organisms and, hence, the population dynamics 
of the species. We have also been able to run our simulations upto 
sufficiently long times ($10^8$, with stationarity achieved at around 
$10^7$) so that the model also accounted for 
macro-evolutionary phenomena like extinctions of species 
as well as speciation that leads not only to emergence of species 
at the existing levels of complexity but also to higher species that 
occupy an altogether new trophic level in the food web. 

From the infinite {\it possible} life forms, we start with one or a 
few, and then let our ecosystem grow in diversity and complexity 
until the limitations of biomass restrict it to hundreds of individuals 
in dozens of species, organized into about five trophic levels.
 
Although our model food web is hierarchical, it is not a tree-like 
structure. The hierarchical architecture helps us in capturing a well 
known fact that in the normal ecosystems the higher is the trophic 
level the fewer are the number of species. It is well known that the 
body size and abundance of a species are strongly correlated to their 
positions as well as to their interactions with other species in the 
food web \cite{cohen93,cohen03}. If we neglect parasites 
and herbivorous insects on trees, then, in general, predators are fewer 
in number and bigger in size as compared to their prey species \cite{cohen03}. 
This is very naturally incorporated in the hierarchical 
food web structure of our model. Let us assume that in the model the 
body size of individual organisms on each level $\ell$ is about $m$ 
times smaller than that on its predator level $\ell-1$. On the other 
hand, the maximum possible populations of organisms, including all the 
nodes, in a level ${\ell}$ is $m$ times that at the level ${\ell}-1$.   
Consequently, the maximum amount of biomass on each level is, 
approximately, the same.  

Since each individual organism appears explicitly in our model, one 
could, at least in principle, assign a genome to each individual and 
describe Darwinian selection which takes place at the level of organisms 
\cite{williams,mayr97}.  Unfortunately, additional ad-hoc assumptions 
would be required to relate the genome with the reproductive success 
\cite{hall,rikvold}.  Instead of introducing an 
ad-hoc mathematical formula to relate genotype with phenotype, we have 
worked directly with phenotype, particularly, quantities that decide 
the reproductive success of the organisms; these quantities are 
$X_{rep}$, $X_{max}$ and $M$. 

From the perspective of self-organization, the new model surpasses all 
cited previous models as not only the characteristic collective properties 
of the species but even the nature of inter-species interactions as well 
as the total number of trophic levels in the food web are determined by 
self-organization of the eco-system.

\noindent {\bf Acknowledgements}

We thank the Supercomputer Center J\"ulich for computer time on their 
CRAY-T3E.  This work is supported by Deutsche Forschungsgemeinschaft 
through a Indo-German joint research project. \\

%%%%%%%%%%%%%%%%%%%%%%%%%%%%%%%%%%%%%%%%%%%%%%%%%%%%%%%%%%%%%%%%%%%%%%
\bibliographystyle{plain}

\end{document}